\documentstyle{l-aa}
\input psfig
\input{s.sty}
\def\simlt{\ \raise -2.truept\hbox{\rlap{\hbox{$\sim$}}\raise5.truept   
\hbox{$<$}\ }}                                                          %
\def\simgt{\ \raise -2.truept\hbox{\rlap{\hbox{$\sim$}}\raise5.truept   %
\hbox{$>$}\ }}                                                          %

\begin{document}

   \thesaurus{03          
              ( 11.03.4;  
                11.09.3;  
                12.03.3;  
                11.01.2;  
                13.25.2)} 

   \title{An X-ray and optical study of the cluster A33}

   \author{S. Colafrancesco
       \inst{1}, C.R. Mullis \inst{2},  A. Wolter \inst{3}, I.M. Gioia
\inst{2,10,11}, 
T. Maccacaro \inst{3}, A. Antonelli \inst{1}, F. Fiore  \inst{1}, J. Kaastra 
\inst{4}, R. Mewe \inst{4}, Y. Rephaeli \inst{5}, R. Fusco-Femiano \inst{6},
V. Antonuccio-Delogu \inst{7}, F. Matteucci \inst{8} and P. Mazzotta \inst{9}}

   \offprints{S. Colafrancesco}

   \institute{Osservatorio Astronomico di Roma 
              via dell'Osservatorio 2, I-00040 Monteporzio, Italy \\
              Email: cola@coma.mporzio.astro.it
\and 
             Institute for Astronomy, University of Hawaii, 
             2680 Woodlawn Drive, Honolulu, HI 96822, USA
\and             
	Osservatorio Astronomico di Brera, Via Brera 26,
        Milano, Italy
\and
        SRON, Sorbonnelaan 2, 3584 CA Utrecht, The Netherlands       
\and
        School of Physics and Astronomy, Tel Aviv University, Israel 69978
\and
        IAS - CNR, Via Fosso del Cavaliere, I00133, Roma, Italy
\and
        Osservatorio Astrofisico di Catania, Via A. Doria,
        Catania, Italy 
\and
        Osservatorio Astronomico di Trieste, Via dell'Osservatorio, Trieste,
        Italy
\and
        Dipartimento di Fisica,  Universit\`a di Roma ``Tor Vergata'',
        Via della Ricerca Scientifica 1, I-00133 Roma, Italy
\and 
    Home institution: Istituto di Radioastronomia del CNR, Via Gobetti 
   101, I-40129, Bologna - Italy
\and 
    Visiting Astronomer at the W. M. Keck Observatory, jointly operated
    by the California Institute of Technology, the University of
    California and the National Aereonautics and Space Administration}

\date{received ; accepted }

\maketitle
\markboth{S. Colafrancesco et al.}{An X-ray and optical study of A33}

\begin{abstract}

We report the first detailed X-ray and optical observations of the 
medium-distant cluster A33 obtained with the Beppo-SAX satellite and with the 
UH 2.2m and Keck II telescopes at Mauna Kea.
The information deduced from X-ray and optical imaging and spectroscopic 
data allowed us to identify the X-ray source 1SAXJ0027.2-1930 as the X-ray 
counterpart of the A33 cluster. 
The faint, $F_{2-10~keV} \approx 2.4 \times 10^{-13} \ergscm2$, X-ray source 
1SAXJ0027.2-1930, $\sim 2$ arcmin away 
from the optical position of the cluster as given in the Abell catalogue,
is identified with the central region of A33.
Based on six cluster galaxy redshifts, we determine the 
redshift of A33, $z=0.2409$; this is lower 
than the value derived by Leir and Van Den Bergh (1977). 
The source X-ray luminosity, $L_{2-10~keV} = 7.7 \times 10^{43} \ergs$, 
and intracluster gas temperature, $T = 2.9$ keV, make this cluster 
interesting for cosmological studies of the cluster 
$L_X-T$ relation at intermediate redshifts.
Two other X-ray sources in the A33 field are identified.
An AGN at z$=$0.2274, and an M-type star, whose emission are blended to
form an extended X-ray emission $\sim 4$ arcmin north of the A33 cluster. 
A third possibly point-like X-ray source detected $\sim 3$ 
arcmin north-west of A33 lies close to a spiral 
galaxy at z$=$0.2863 and to an elliptical galaxy at the same redshift 
as the cluster.


\end{abstract}

\section{Introduction}

A33 is a medium-distant Abell cluster of galaxies with
very few and sparse information in both the X-ray and the optical bands.
This cluster was claimed to have been detected by the 
HEAO1-A1 all sky survey (Johnson et al. 1983, Kowalski et al. 1984)
with a count rate of $3.77 \pm 0.47$ counts  cm$^{-2}$ s$^{-1}$ 
in the $2-6$ keV energy band.
Its luminosity was estimated, with large uncertainties, 
to be $L_{2-6~keV} \approx 2.34 \times 10^{45}$ erg s$^{-1}$.

A33 was also observed with the GINGA LAC detector from 
December 9 to December 10, 1988 (Arnaud \ea 1991),
but no X-ray emission was found at the optical position of 
the cluster.  From such a non-imaging observation, Arnaud \ea (1991) 
were able to put an upper limit on the
luminosity of A33, $L_{2-10~keV} < 6 \times 10^{44}$ erg s$^{-1}$, 
assuming a temperature $T = 8.4$ keV.
The value of the X-ray luminosity derived from GINGA data
is inconsistent with the one derived from the 
HEAO1-A1 observation (note, however, that A33 lies at the edge of 
the error box for the position of the HEAO1 source).

The source 1RXSJ002709.5-192616 in the ROSAT Bright Source Catalog 
(BSC: Voges et al., 1996), 
at coordinates $\alpha^x_{2000}= 00^h~ 27^m~ 09.50^s$  and
$\delta^x_{2000}=-19^o~ 26'~ 16"$, has been observed for 
$317$ sec with a count rate of $0.062 \pm 0.017$ cts/s. This source 
has $19.6$ net counts in the $0.1-2.4$ keV energy band
corresponding to a flux 
$F_{0.2-2.4}=(9.3 \pm 2.6) \times 10^{-13}$ erg s$^{-1}$ cm$^{-2}$
(assuming a nominal conversion factor of 
$1.5 \times 10^{-11}$ erg cm$^{-2}$ s$^{-1}$ cts$^{-1}$) and does 
not appear to be extended. This source is unrelated to the cluster
and most probably associated with an AGN which is only $5.4''$ away
(see Table 1, source 1SAXJ0027.1-1926, and Table 2, source A).

In the optical band there is no detailed 
information except from that derived from the extensive study
of  Leir \& Van Den Bergh (1977), who 
classified A33 as a distance class $D=6$, richness $R=1$, 
Bautz-Morgan-class-III cluster. In the Abell (1958) catalog, A33 
has $69$ galaxies which lie within one Abell radius ($2.7 z^{-1}$ arcmin) 
and which are not more than 2 mag fainter than the third brightest galaxy.
Its photometrically estimated redshift, $z = 0.28$, 
was derived by Leir \& Van Den Bergh (1977) from the cluster optical
diameter and the magnitude of the brightest and tenth-brightest
cluster galaxies.

In this paper we present a new X-ray observation of A33 
obtained with Beppo-SAX. 
This observation enables us to derive detailed information on
the X-ray source, on its morphology and thermal properties.
The complex appearance of the  X-ray emission in the field of 
A33 prompted us to obtain optical images and spectroscopic 
information for several objects in the field.

The plan of the paper is the following.
In Section 2 we present the basic information on the Beppo-SAX 
observation and  data reduction. In Section 3 we describe 
the optical data and in Section 4 we discuss the X-ray spectroscopy
of the various sources in the A33 field. 
We summarize our results for A33 and discuss their implications in
Section 5.

Throughout the paper $H_0$=50 km sec$^{-1}$ Mpc$^{-1} $
and $\Omega_0=1$ are used unless otherwise noted.
{\footnotesize
\begin{table*}[htbp]
\begin{center}
\caption{LECS ($0.1-2$ keV) and MECS ($2-10$ keV) count rates}
\label{tab_1}
\begin{tabular}{ccccccc}
\hline \hline
Source & $\alpha_{2000}$ & $\delta_{2000}$ & $t_{exp}$  &
Count rate  &  Count rate & $R_{extr}$ \\ \hline
       &$(^h~ ^m~ ^s)$ &$(^o~'~'')$  & ($s$)  & ($10^{-3} s^{-1}$) &  ($10^{-3} s^{-1}$) 
& arcmin \\ 
       & &  &        & LECS  & MECS  & \\
\hline\hline
1SAXJ0027.1-1926   & $00~ 27~ 08$  &
$-19~ 26~ 38$ & $77609$  & $7.6 \pm 0.7$  & $7.8 \pm 0.6$ & 2  \\
1SAXJ0027.2-1930  & $00~ 27~ 12$  &
$-19~ 30~ 32$ & $77609$  &  --   &  $1.76 \pm 0.23$  & 2       \\
1SAXJ0027.0-1928  & $00~ 27~ 01$  &
$-19~ 28~ 30$ & $77609$    &  --    &  $ 1.14 \pm 0.19$   & 1     \\
\hline \hline
\end{tabular}
\end{center}
\end{table*}
}

\section{Beppo-SAX Observation}

The A33 field was observed with the Narrow Field Instruments (NFI)
of the Beppo-SAX satellite from November 23$^{\rm th}$ to 
25$^{\rm th}$, 1996. 
The total effective exposure time is
$t_{exp}= 3.8417 \times 10^4$ s 
for the LECS instrument and $t_{exp}= 7.7610 \times 10^4$ s 
for the MECS instrument (see \eg Boella et al. 1997a and 1997b for a technical
description of the Beppo-SAX mission and instrumentation).   
\begin{figure*}
\begin{center}
\psfig{figure=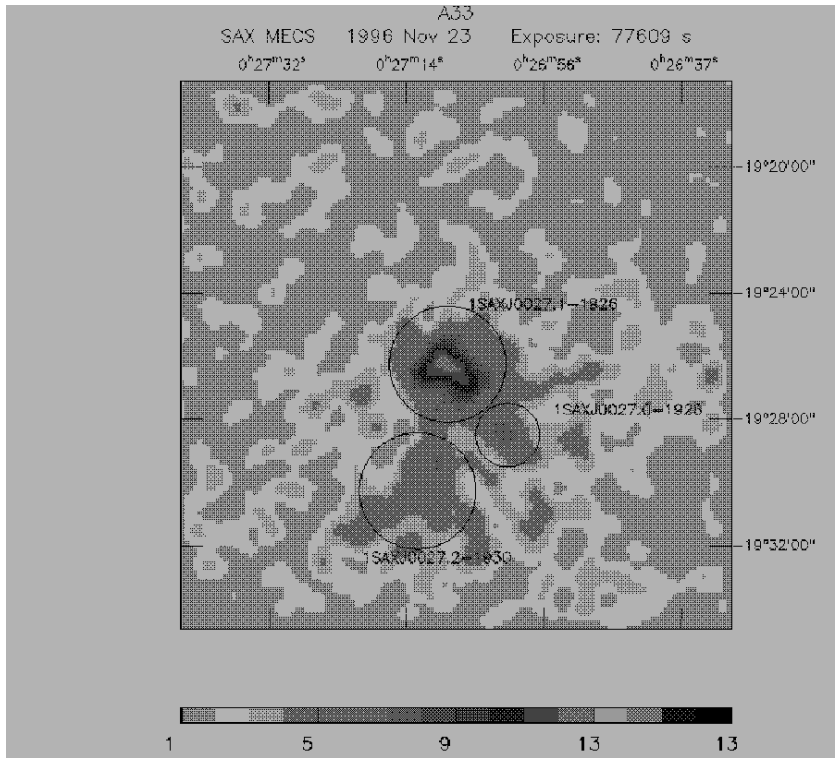,height=10.cm,width=11.cm,angle=0}
\end{center}
\caption{\footnotesize {The Beppo-SAX image of A33 in the $2-10$ keV 
energy band. The three different components of the emission are labeled
according to the text. The circles indicate the extraction area for each
X-ray source. Note that 1SAXJ0027.2-1930 has also a diffuse, low-surface 
brightness distribution which appears to be extended in the
southern part of the image. The image has been deconvolved with a wavelet 
transform using a smoothing length  of 3.5 pixels (1 pixel = 8 arcsecs).
North is up and East to the left.
}}
\label{figure:fig_1}
\end{figure*}

Data preparation and linearization was performed using the 
SAXDAS v.1.3 package under the FTOOLS environment. 
The imaging analysis was performed using the 
XIMAGE package (Giommi \etal 1991).
The extraction of the source and background spectra was done 
within the XSELECT package.
The spectral analysis was performed using XSPEC v.9.0.

The only previous claimed X-ray detection of A33 was done with 
the HEAO1 satellite (Johnson et al. 1983; Kowalski et al. 1984). 
Due to the large error box of the HEAO1 detectors, 
the coordinates of the X-ray source were associated with the optical
coordinates of the A33 cluster. Thus the 
Beppo-SAX observation was centered on the optical coordinates
$\alpha^o_{2000}=$ $00^{h}26^{m}52.7^{s}$ and 
$\delta^o_{2000}= -19^{o}32'29"$.
The MECS $2-10$ keV X-ray image of the field is shown in 
Fig.\ref{figure:fig_1}, where three different subsystems are 
evident: a bright and apparently extended  source, 
1SAXJ0027.1-1926, an extended but smaller source, 1SAXJ0027.2-1930, 
located to the south of the brightest source and an apparently 
point-like source, 1SAXJ0027.0-1928, located to the west.
Positions, count rates and extraction region radii, $R_{extr}$, 
are listed in Table \ref{tab_1}.
The sources have sufficient count rates to be detected
individually at more than $4$ sigma level by the MECS instruments.
The poorer spatial resolution of the LECS instead allows only to determine the
count rate of the brightest source 1SAXJ0027.1-1926.
In the following we describe the spatial structure of each 
source detected in the A33 field as derived from the MECS data.

The MECS PSF is $\approx 1$ arcmin Half Energy Width, and this spatial resolution
allows us to detect the sources 1SAXJ0027.1-1926 and 1SAXJ0027.2-1930 as extended in the
MECS image of Fig.1.

\begin{figure}
\psfig{figure=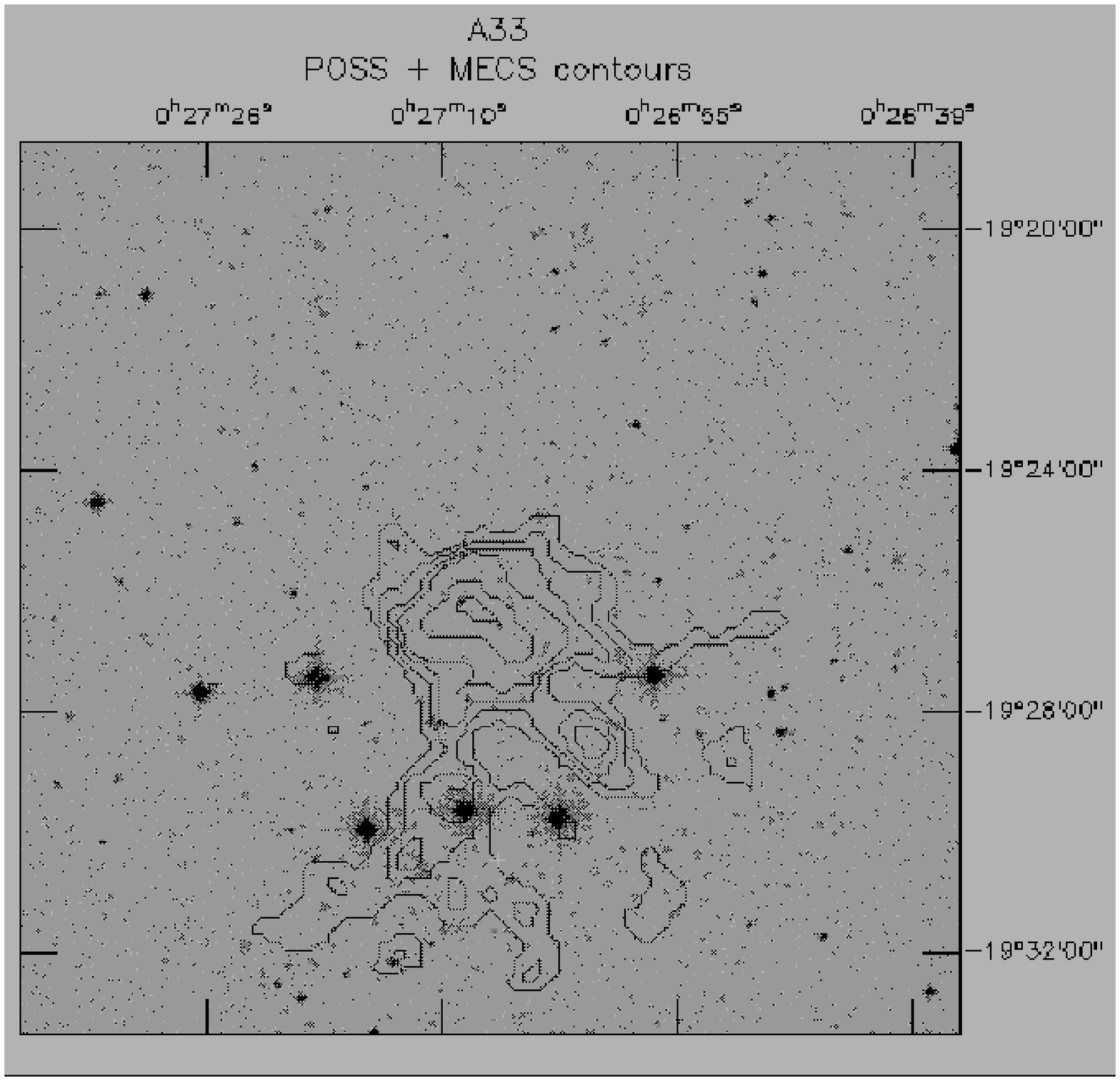,height=9.cm,width=9.cm,angle=-90.}
\caption{\footnotesize {The optical image of A33 taken from the
POSS II plate and the X-ray contours of the Beppo-SAX image obtained with the 
MECS detector in the $2-10$ keV energy band. Contours are taken from the image 
shown in Fig.1 and are logarithmically spaced.
The image has been deconvolved with a wavelet transform using
a smoothing length  of 3.5 pixels (1 pixel = 8 arcsecs).
The white cross indicates the position of A33 from the Abell catalogue.
The first X-ray contour is at $3 \sigma$ from the background level.
North is up and East to the left.
}}
\label{figure:fig_2}
\end{figure}

The source 1SAXJ0027.1-1926  has an extension of $\sim 2$ arcmin (radius).
As discussed in Sections 3 1nd 4, this source is most probably the result
of the blending of two point-like sources not resolved by the MECS PSF.
The X-ray MECS image contours superposed onto the POSS II image of
the field plotted in Fig.\ref{figure:fig_2}
show that there is no clear galaxy excess
associated to the X-ray source 1SAXJ0027.1-1926. 

The source 1SAXJ0027.2-1930, located $\sim 4.5$ arcmin 
south of the brightest source (see Fig. 1), has an extension of 
$\simgt 1.5$  arcmin radius. Using a $\beta$-model with values $\beta=0.75$ 
and $r_c=260$ kpc ($H_0=50, \Omega_0=1$) chosen as representative of such 
low luminosity objects, and convolved with the MECS PSF
we find a central density of $\approx 3.9 \cdot 10^{-3}$ cm$^{-3}$.
Moreover, an extended, low surface brightness  X-ray emission 
is visible in the southern part of the image (see Fig. 1 and Fig. 2).
Such a low surface brightness source  extends for a few arcminutes
at levels of $\sim 10^{-4}$ cts s$^{-1}$ cm$^{-2}$ arcmin$^{-2}$.
The extended source  1SAXJ0027.2-1930 is associated with A33 as shown
in the POSS II image of the field (see Fig.\ref{figure:fig_2} and 
Section 3).

The third source 1SAXJ0027.0-1928, located $\sim 4$ arcmin south-west 
of the brightest source, has a  point-like appearance.
Two faint objects in the POSS II are positionally consistent with
1SAXJ0027.0-1928.

\section{Optical Imaging and Spectroscopy}
\def\arcdeg{\hbox{$^\circ$}}
\def\arcsec{\ifmmode^{\prime\prime}\;\else$^{\prime\prime}\;$\fi}
\def\arcmin{\hbox{$^\prime$}}
{\footnotesize
\begin{table*}[htbp]
\begin{center}
\caption{Optical results}
\label{tab_o}
\begin{tabular}{ccccl}
\hline \hline
Name & $\alpha_{2000}$ & $\delta_{2000}$ & $z$ & Identification and  Comments \\ 
     & $(^h~ ^m~ ^s)$  &  $(^o~'~'')$    &     &                              \\
\hline\hline
 A   & 00 27 09.8  & $-$19 26 12.6 & $0.2274 \pm 0.0006$ & AGN 
  ([OII], [OIII], [Ne III], broad Balmer) \\
 B   & 00 27 07.3  & $-$19 26 36.4 &  &  M star    \\
 C   & 00 27 00.5  & $-$19 28 56.5 & $0.2420 \pm 0.0005$ & galaxy (G-band, H$\beta$, MgIb, NaId) \\
 D   & 00 26 59.5  & $-$19 28 18.6 & $0.2863 \pm 0.0015$ & galaxy (H+K, G-band, 
H$\beta$, MgIb) \\
g1   & 00 27 12.3  & $-$19 30 45.5 & $0.2406 \pm 0.0008$ & galaxy (H+K, G-band, 
H$\beta$, MgIb, NaId) \\
g2   & 00 27 12.6 & $-$19 30 43.7  & $0.2380 \pm 0.0012$ & galaxy (CaII-break, G-band, H$\beta$, MgIb, NaId) \\
g3   & 00 27 12.5 & $-$19 30 40.1  & $0.2395 \pm 0.0017$ & galaxy (CaII-break, 
G-band, H$\beta$, MgIb, NaId) \\
g4   & 00 27 13.1 & $-$19 30 29.4  & $0.2445 \pm 0.0004$ & galaxy (H+K, G-band, 
H$\beta$, MgIb, NaId) \\
g5   & 00 27 13.0 & $-$19 30 25.6  & $0.2406 \pm 0.0005$ & galaxy (H+K, G-band, 
H$\beta$, MgIb, NaId) \\
\hline \hline
\end{tabular}
\end{center}
\end{table*}
}

\begin{figure}
\begin{center}
{\vbox{
\psfig{figure=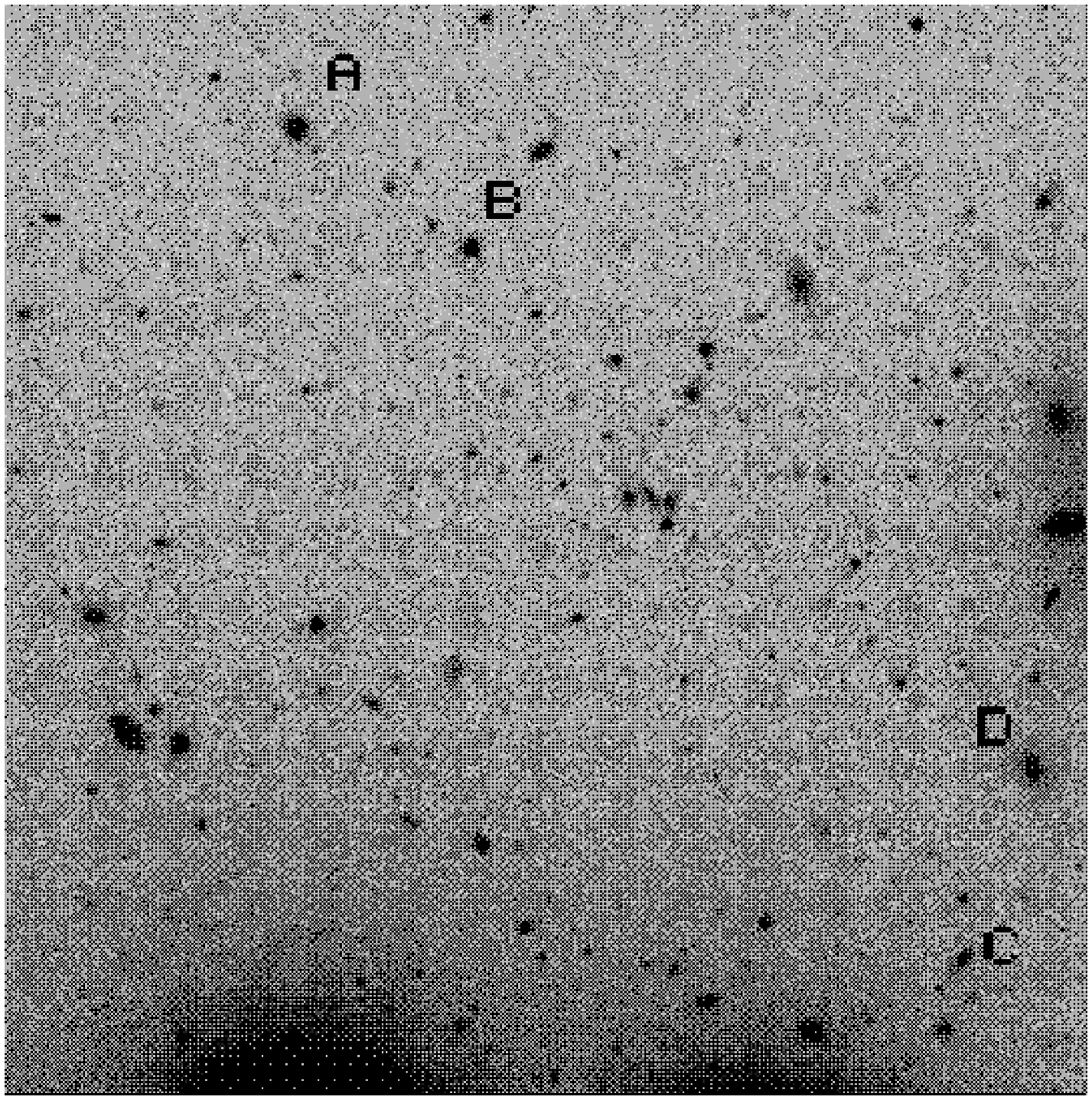,height=7.5cm,width=7.5cm,angle=0.}
\psfig{figure=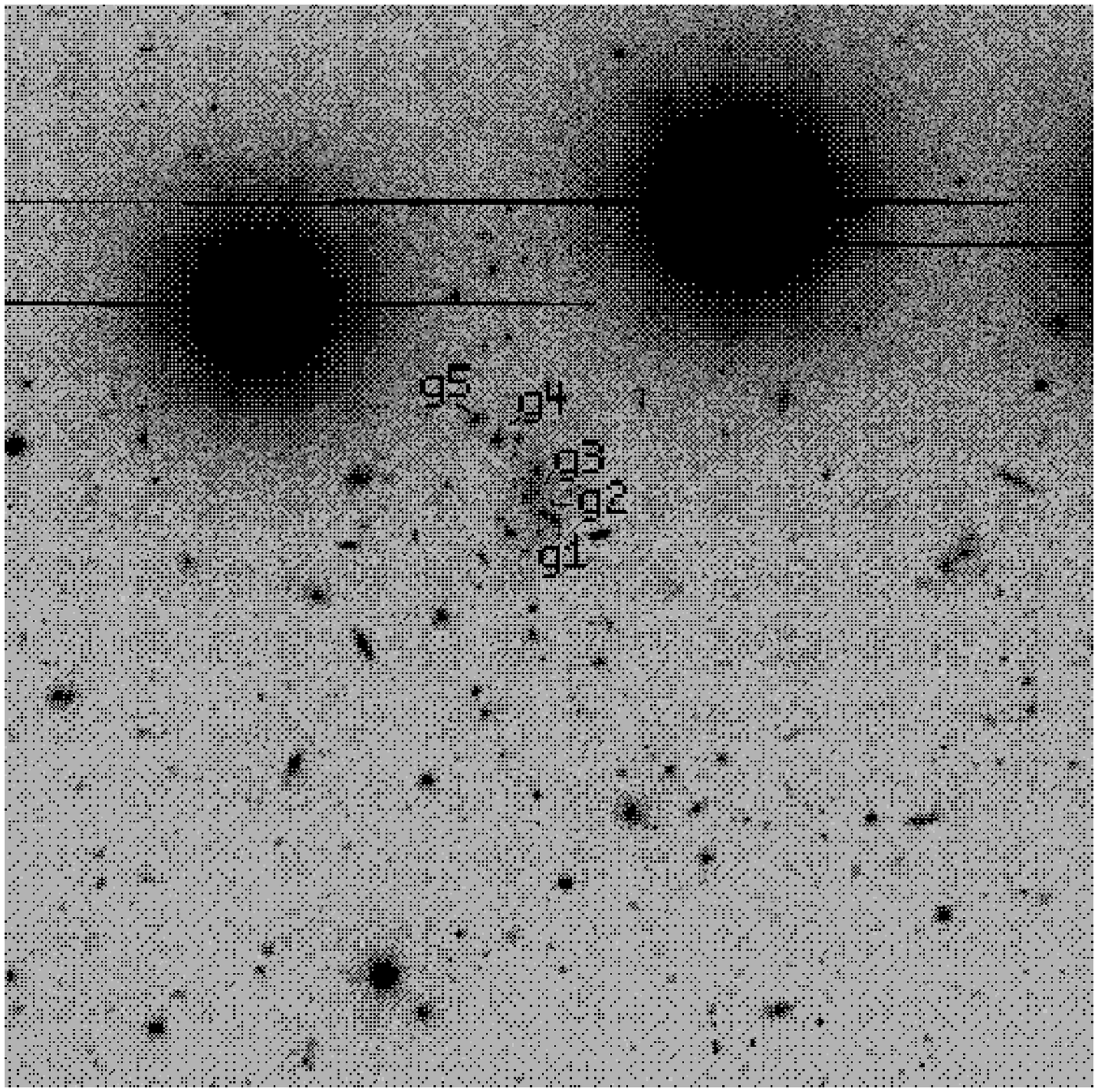,height=7.5cm,width=7.5cm,angle=0.}
}}
\end{center}
\caption{\footnotesize {The two images are 1024x1024 (3.7x3.7 arcmin) 
subarrays 
extracted from two B-band exposures taken  at the Keck II telescope. 
The image to the top shows the field around 1SAXJ0027.1-1926 and 
1SAXJ0027.0-1928 and the image to the botton shows the field around 
1SAXJ0027.2-1930. 
North is up and East to the left.
}}
\label{figure:fig_10}
\end{figure}

Due to the lack of detailed optical information in the literature 
for A33, we took I and B images of the cluster region on 
November 23 and 24 1997 at the Keck II telescope. 
The images were  obtained using the Low-Resolution and Imaging 
Spectrograph (LRIS)  (Oke et al. 1995) in imaging mode, resulting in 
a scale of 0.215\arcsec pixel$^{-1}$ and a field of view of 
6\arcmin$\times$7.\arcmin3. The I (B) images were taken in
0.4\arcsec$-$~0.5\arcsec seeing on the first night and consist of 
3$\times$300s (4$\times$120s) dithered exposures 
centered at $\alpha$=00$^{h}$27$^{m}$10.$^{s}$5 and
$\delta=-19\arcdeg29\arcmin18\arcsec$ (J2000), the southern region 
of the X-ray emission complex. On the second night (0.8\arcsec seeing)
we took 2$\times$120s I (2$\times$300s B) exposures centered at 
$\alpha$=00$^{h}$27$^{m}$09.$^{s}$8 and 
$\delta=-19\arcdeg26\arcmin12.\arcsec4$ (J2000), 
the northern region of the X-ray emission system. 
The optical position of A33 (Fig.\ref{figure:fig_2}) is close to 
an open stellar cluster. 
Fig.\ref{figure:fig_10} 
shows the B images for both North (Fig.\ref{figure:fig_10}a) and  
South (Fig.\ref{figure:fig_10}b) regions. 
No excess of galaxies is present in the  northern region at the 
position of 1SAXJ0027.1-1926 (Fig.\ref{figure:fig_10}a), while 
Fig.\ref{figure:fig_10}b reveals an overdensity of galaxies
in the region of the X-ray source 1SAXJ0027.2-1930.

Spectroscopic observations for several objects in the
field were carried out on August 16, 17
and 19, 1998, with the Wide Field Grism Spectrograph and the 
Tek2048$\times$2048 CCD attached to the University of Hawaii 2.2m 
telescope on Mauna Kea.  We used the 420 l/mm grating which provided a 
$\sim$3990-9900 \AA\ coverage and a pixel size of 3.6 \AA/pix, and a 
long-slit of 2.4$''$ which gives a low spectral resolution of about 24 \AA.
For the reduction of the data we have used the IRAF package (Tody, 1993).
In the region of the northern X-ray emission we identified 2 objects
labeled as A and B in Fig.\ref{figure:fig_10}a. 

In the region to the west, where the X-ray source 1SAXJ0027.0-1928 is 
present, we found two galaxies labeled C and D in the 
above mentioned figure. In the region of the southern X-ray 
emission we obtained spectra for five galaxies which turned
out to be members of the cluster. These galaxies are labeled g1 through 
g5 in Fig.\ref{figure:fig_10}b. Table \ref{tab_o} gives the results of the observations:

Based on our imaging and spectroscopic results, we conclude that a
blend of the AGN (A) and M-type star (B) X-ray emissions contribute
to the extended source 1SAXJ0027.1-1926 to the north.
The Abell cluster A33 is the source of the southern X-ray emission
1SAXJ0027.2-1930, while the identification of the source of the 
western X-ray emission, 1SAXJ0027.0-1928, remains unknown.
The two galaxies for which we measured the spectra,
and which are the two brightest optical sources in the region,
might be responsible for part of the emission of
1SAXJ0027.0-1928, but we need spectroscopic data for more objects to 
help in the identification. 
One of the sources (C) is consistent with being part of A33.
From the six cluster members listed in 
Table \ref{tab_o} we obtain for A33 an  average $<z>=$0.2409$\pm$0.0009, 
and a very tentative velocity dispersion, given the few cluster galaxies,
$\sigma_{los}=$472$^{+295}_{-148}$ km s$^{-1}$.
This estimate includes the $1+z$ correction.

\begin{figure}
\psfig{figure=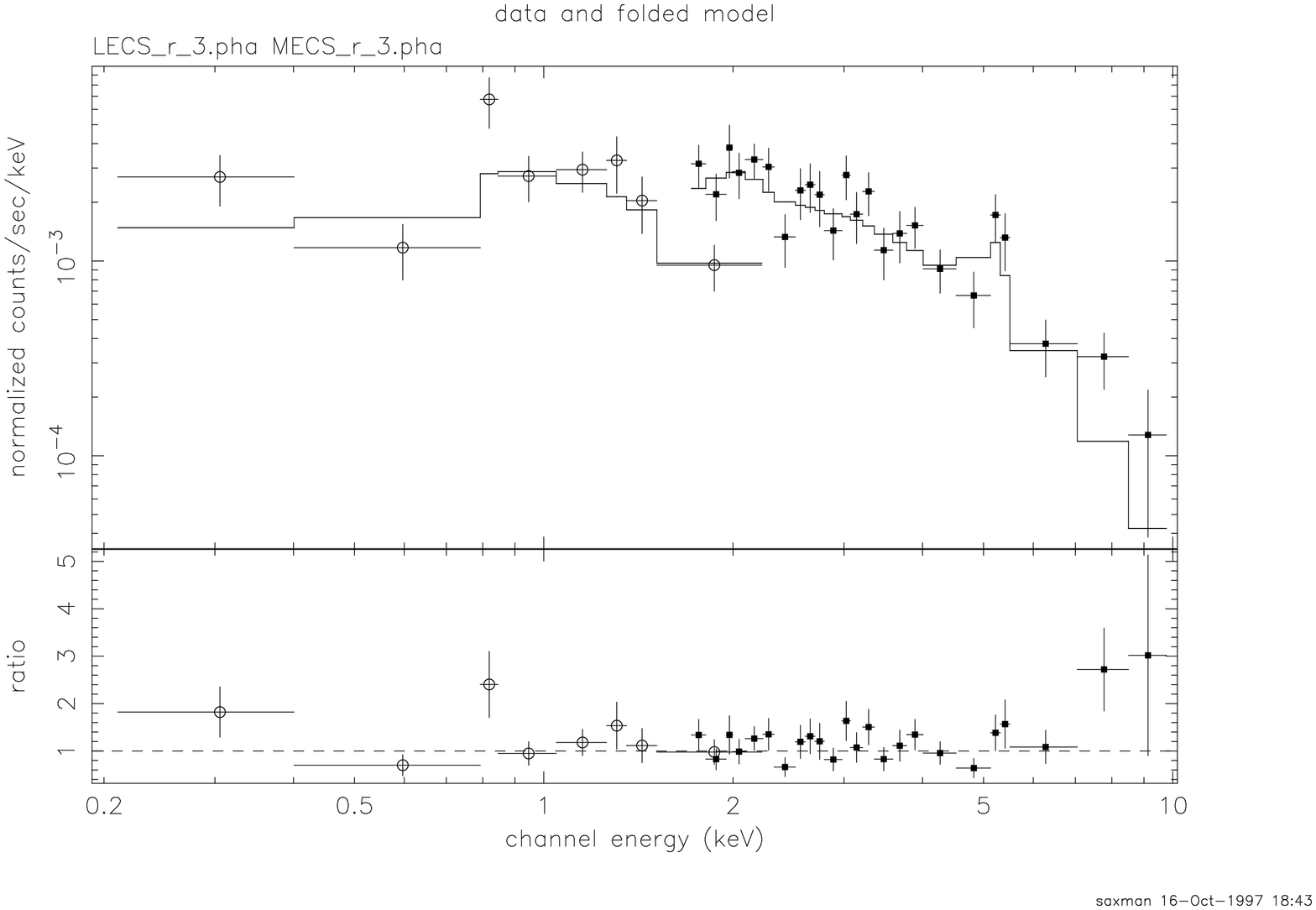,height=9.cm,width=8.5cm,angle=-90.}
\caption{\footnotesize {The combined LECS and MECS spectrum  
of the source 1SAXJ0027.1-1926 extracted from a $2$ arcmin radius 
regions. The spectrum shown in figure has been rebinned so that the 
significance of each bin is at least 3$\sigma$. 
The best fit model is a MEKAL thermal model (see text for details).
The spectrum has been further rebinned using XSPEC for graphical purposes.  
}}
\label{figure:fig_5}
\end{figure}

\section{X-Ray Spectroscopy}
The Beppo-SAX concentrator/spectrometer system consists of four 
separated concentrator mirrors, three of  them covering the 
$1.6- 10$ keV range (Medium Energy Concentrator Spectrometer, or 
MECS)  and the  fourth extending to lower energies
down to $0.1$ keV (Low Energy Concentrator  Spectrometer, or LECS). 
The concentrators are designed to have a large 
effective area around the iron K$\alpha$ line complex:
$150$  and $50$ cm$^2$  for  MECS and LECS,  at $6$ keV. 
Also, Beppo-SAX is able to provide spatially resolved spectra: its energy 
and angular resolution are  $\Delta E/E=8\%$ at $6$ keV and 
$\theta_{FWHM} \approx 40''$, respectively. 

{\footnotesize
\begin{table*}[htbp]
\begin{center}
\caption{1SAXJ0027.1-1926}
\label{tab_2}
\begin{tabular}{ccccccc}
\hline \hline
Model & pho. index & $z$ & $T$  &  bins & $\chi^2$ & $\chi^2_{red}$\\ \hline
       &            &      & keV      &       &         \\ 
\hline\hline
RS     &    --      & $0.245 \pm 0.023$ & $3.99^{+0.95(+1.83)}_{-0.77 (-0.98)}$
 &  61   & 56.12 & 0.97 \\
MEKAL  &    --      & $0.245 \pm 0.024$ & $3.90^{+0.99(+1.88)}_{-0.66 (-0.97)}$ &  61   & 55.98 & 0.97      \\
PL     & $2.05^{+0.17(+0.41)}_{-0.14 (-0.18)}$     &  --  & --     
&  61   & 62.55 & 1.06  \\
\hline \hline
\end{tabular}
\end{center}
\end{table*}
}

In order to obtain the emission weighted spectral information 
of the three main sources in the A33 field,
we have extracted the photons from 
circular regions drawn around each source 
(see Fig.\ref{figure:fig_1}). The extraction radius, smaller than the 
suggested $4$ arcmin radius region since the sources are separated by a small
angular distance, might introduce a systematic uncertainty.
We have used the appropriate Ancillary Response File to correct for this
effect. We fitted the source spectra using both a Raymond-Smith code
(1977;  hereafter RS) or a MEKAL code (Mewe, Kaastra \& Liedahl 1995) 
to model the thermal intracluster gas emissivity and
a simple absorbed power-law, non-thermal model.
Background spectra have been extracted from library blank-sky images in 
the same circular regions as the sources.
\begin{figure}
{\vbox{
\psfig{figure=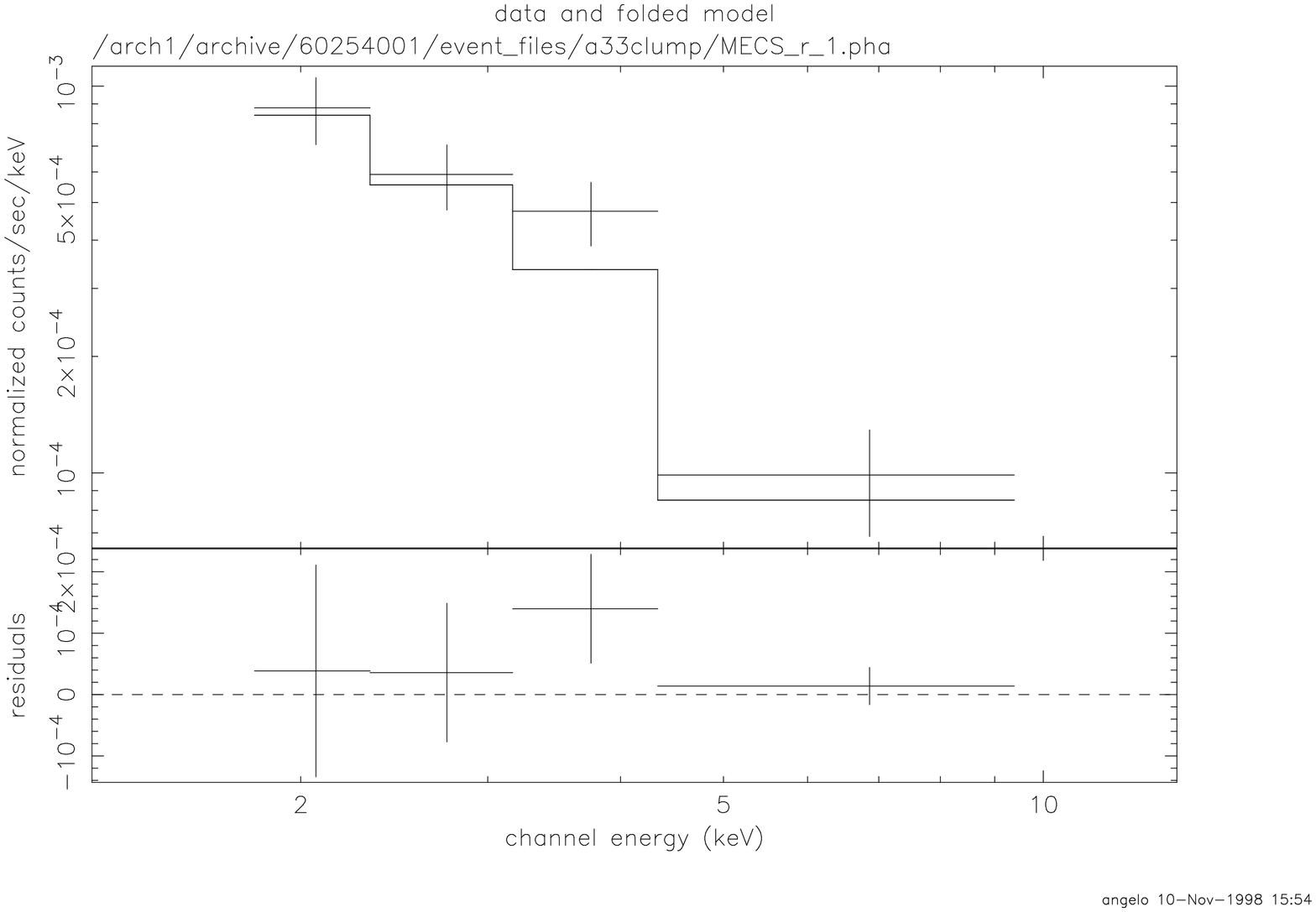,height=9.cm,width=8.5cm,angle=-90.}
\psfig{figure=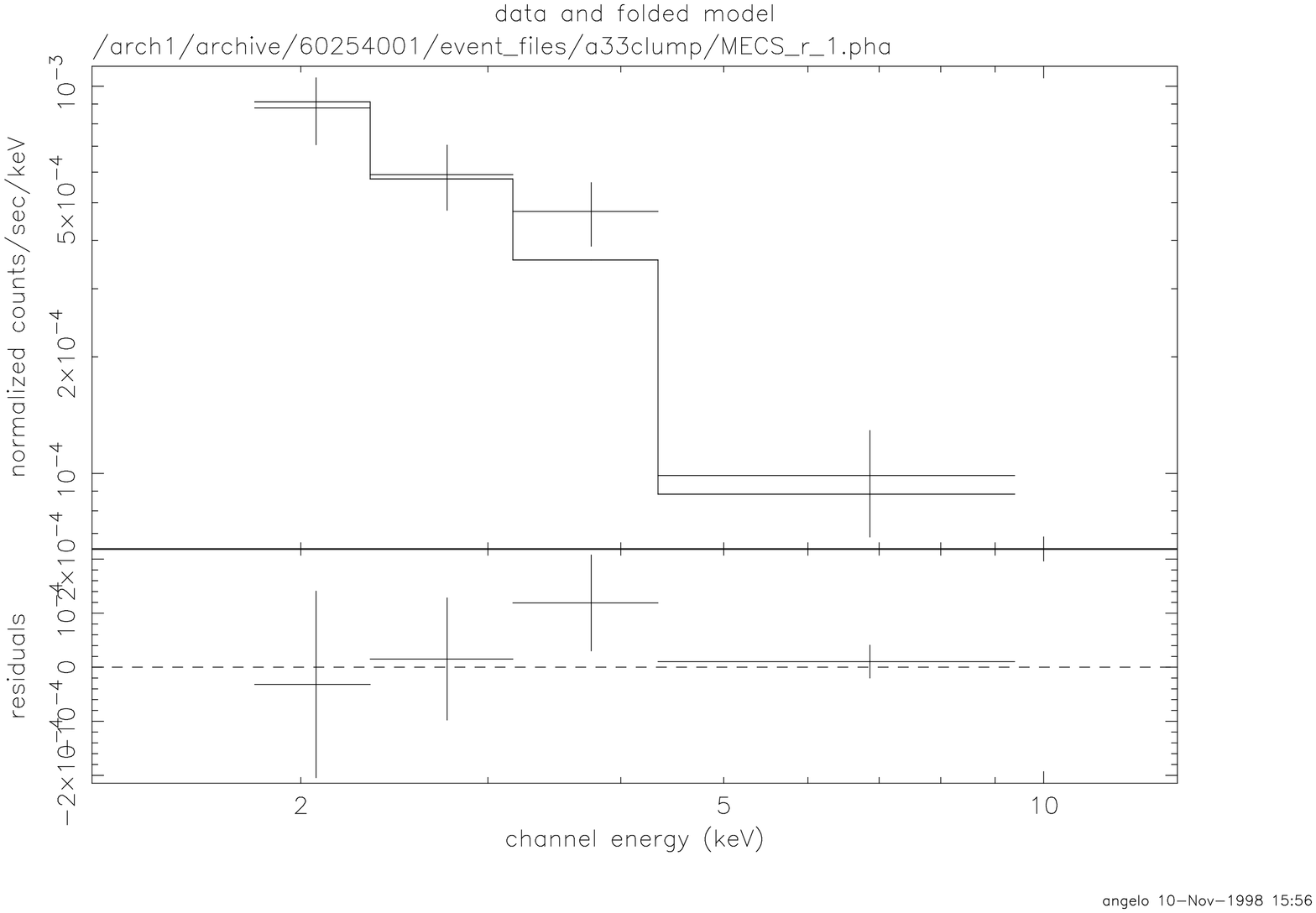,height=9.cm,width=8.5cm,angle=-90.}
}}
\caption{\footnotesize {The SAX MECS spectra of 1SAXJ0027.2-1930  
fitted with a 
thermal MEKAL model (upper panel) and with an absorbed power-law 
model (lower panel). Details of the spectral analysis are given in
Table 4.
The spectrum has been further rebinned using XSPEC for graphical purposes.  
}}
\label{figure:fig_7}
\end{figure}

\vskip 0.3truecm
\noindent
{\it a) 1SAXJ0027.1-1926}

\noindent
The spectrum of the brightest source in the field was extracted, both 
for the LECS and the MECS instruments, from a circular region of $2$ 
arcmin radius centered on the X-ray position of Table 1.
The combined LECS-MECS spectrum is shown in Fig.\ref{figure:fig_5}: 
we do not observe any low energy excess absorption in the spectrum,
thus we keep $N_H$ fixed at the galactic value of 1.86 $\times 10^{20}
cm^{-2}$ (Dickey \& Lockmann, 1990) relative to the source position.

The best fit spectral parameters for the MECS spectrum are listed  in 
Table \ref{tab_2}  together with their uncertainties at $68.3 \%$ (and $90 \%$ 
in parentheses) confidence level.
We use $605$ source photons in this spectral fit.

Within $2$ arcmin from its center, the source has a flux of
$F_{2-10 keV} = (4.20\pm0.32) \times 10^{-13}$ erg s$^{-1}$ cm$^{-2}$, evaluated using 
the MEKAL best fit parameters. 
The other models give similar fluxes.
This flux is also consistent, within the errors, with the flux of the X-ray source
1RXSJ002709.5-192616 in the ROSAT band.

The optical magnitude of the M-star if $m_V \approx 19$.
Assuming that the X-ray flux of the M-star contributes to $50 \%$ of the 
total flux of 1SAXJ0027.1-1926, we obtain $F_{2-10}/F_V \approx 2.$ in the 
$2-10$ keV band and  $F_{0.3-3.5}/F_V \approx 6$ in the $0.3-3.5$ energy 
band (assuming a thermal emission at $T=1$ keV).
This ratio is almost one order  of magnitude higher than the values of 
$F_{0.3-3.5}/F_V$ for X-ray selected stars in the EMSS 
(see Fig.1 in Maccacaro et al. 1988). 
This means that the contribution of the M-star to the 
X-ray flux of 1SAXJ0027.1-1926 should be $\simlt 8 \%$ to be 
consistent with the values of $F_X/F_V$ for normal stars.
If this is the case, then more than half 
of the X-ray emission of  1SAXJ0027.1-1926 is due to the AGN (listed 
as A in Table \ref{tab_o}) at $z=0.2274$ with a luminosity
$L_{2-10~keV} \simlt 4.5 \times 10^{43}$ erg s$^{-1}$.
Otherwise, the source 1SAXJ0027.1-1926 should result from the blend of
the AGN and of a different unknown X-ray source.

{\footnotesize
\begin{table*}[htbp]
\begin{center}
\caption{1SAXJ0027.2-1930}
\label{tab_3}
\begin{tabular}{ccccccc}
\hline \hline
Model & pho. index & $z$ & $T$  &  bins & $\chi^2$ &  $\chi^2_{red}$  \\ \hline
       &            &      & keV      &       &         \\ 
\hline\hline
RS     &    --      & $0.2409$ & $2.91^{+1.25 (+2.44)}_{-0.54 (-0.83)}$   
&  19   & 20.43 & 1.28 \\
MEKAL  &    --      & $0.2409$ & $2.88^{+1.23 (+2.46)}_{-0.55 (-0.83)}$ &  
19   & 20.45 & 1.28      \\
PL     & $2.67^{+0.41 (+0.67)}_{-0.37 (-0.57)}$     &  --  & --   
& 19   &  22.05 & 1.30      \\
\hline \hline
\end{tabular}
\end{center}
\end{table*}
}

\vskip 0.3truecm
\noindent
{\it b) 1SAXJ0027.2-1930}

\noindent
The average spectrum of 1SAXJ0027.2-1930 was extracted from a circular region of
$2$ arcmin radius centered on the X-ray position of Table 1 (see also Fig.5).
In this region there is a clear excess of galaxies (see Fig. 3) which is
$\sim 1.5$ arcmin away from the Abell catalog position of A33.
Fitting the spectrum (which contains $140$ source photons) 
with a RS thermal model with temperature, abundance and 
redshift as free parameters, the fit gives $\chi^2_{red} = 1.22$.  
Fixing the value of $N_H$ to the galactic value (1.86$\times 10^{20}$ cm$^{-2}$)
we obtain an average temperature $T = 3.1\pm0.9$ keV and a redshift 
$z_X = 0.72 \pm 0.04$. The abundance is only marginally constrained at 
$Fe/H = 0.98 \pm 0.71$ of the solar value. However, the fit results are 
mainly due to a marginally significant spectral feature at $E \sim 4$ keV.

Therefore, we fixed the redshift of the X-ray source at $z = 0.2409$, 
as measured from the optical spectra (see Section 3), and
we fitted the spectrum again, fixing the abundance to a value 
Fe/H = 0.3 solar. The results of the fit are shown in Table 4.
Uncertainties in the temperature of 1SAXJ0027.2-1930 are given at $68.3 \%$ 
(and $90 \%$ in parentheses) confidence level.
The low count rate of the source does not allow a more accurate 
description of the X-ray emission.

Assuming the MEKAL best fit parameters we obtain an integrated flux 
of $F_{2-10 keV} = (2.4\pm0.3) \times 10^{-13} \ergscm2$ 
in the 2 arcmin radius extraction region (which corresponds to a
linear size of $\approx 1 ~h^{-1}_{50}$ Mpc). The other models give 
consistent fluxes. At the redshift of the cluster this 
flux corresponds to a luminosity  $L_{2-10 ~keV} = (7.7\pm0.9) 
\times 10^{43} h^{-2}_{50}$ erg s$^{-1}$ and to a bolometric luminosity
$L_{bol}=(2.2\pm0.3)\times 10^{44} h^{-2}_{50}$ erg s$^{-1}$.
\begin{figure}
\psfig{figure=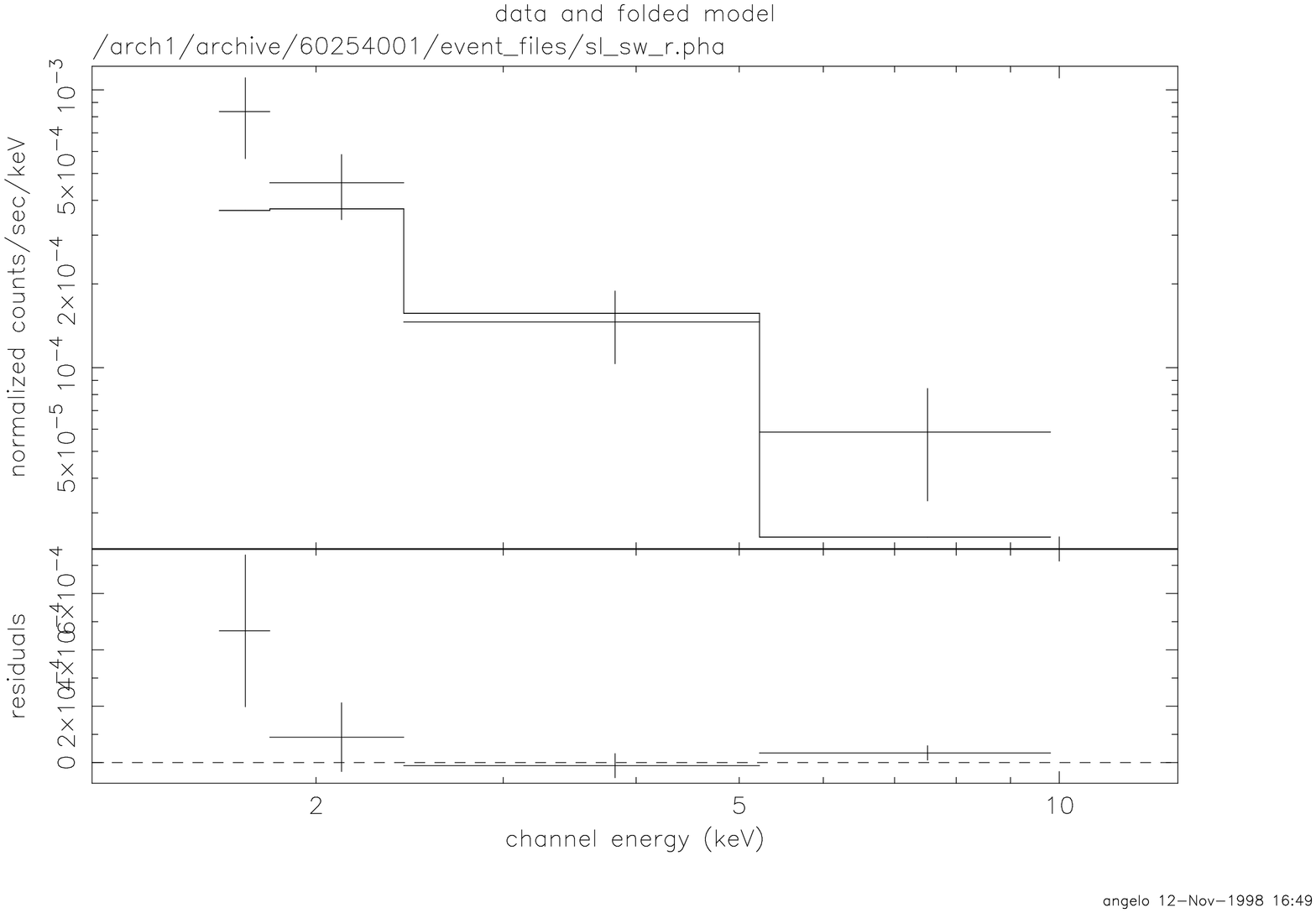,height=9.cm,width=8.5cm,angle=-90.}
\caption{\footnotesize {The SAX MECS spectrum of 1SAXJ0027.0-1928 
fitted with a non-thermal power-law model (see Table 4 for details).
The spectrum has been further rebinned using XSPEC for graphical purposes.  
}}
\label{figure:fig_9}
\end{figure}
{\footnotesize
\begin{table*}[htbp]
\begin{center}
\caption{1SAXJ0027.0-1928}
\label{tab_4}
\begin{tabular}{ccccccc}
\hline \hline
Model & pho. index & $z$ & $T$  &  bins & $\chi^2$ & $\chi^2_{red}$  \\ \hline
       &            &      & keV      &       &         \\ \hline
\hline
RS     &    --      & $0.36 \pm 0.09$ & $2.56\pm 0.86$   &  13   & 14.70 & 1.47     \\
MEKAL  &    --      & $0.35 \pm 0.11$ & $2.45\pm 0.85$   &  13   & 14.79 & 1.48     \\
PL     & $2.63 \pm 0.60$     &  --  & --                 &  13   & 14.98 & 1.36     \\
\hline \hline
\end{tabular}
\end{center}
\end{table*}
}

\vskip 0.3truecm
\noindent
{\it c) 1SAXJ0027.0-1928}

\noindent
We extracted the spectrum of  1SAXJ0027.0-1928 from a circular region of $1$
arcmin radius centered on the X-ray position of Table 1 (see 
Fig.\ref{figure:fig_9}). Results of the fit are shown in Table 5
(uncertainties on the best fit values are given here at 68.3 $\%$ confidence
level. Note that the spectrum of this source contains $90$ source photons).
Assuming an absorbed power--law non--thermal model, we derived a flux of
$F_{2-10 keV} = (4.74\pm0.8) \times 10^{-14} \ergscm2$.
There are two galaxies, a spiral (C) at the same $z$ of A33, 
and an elliptical (D) in the 
region for which we took an optical spectrum. 

The identification of the source is not certain at the moment.
Assuming that the galaxy D at $z = 0.2863$ is the X-ray emitter, its 
X-ray luminosity would be $L_{2-10~keV} = 1.9 \times 10^{43}$ erg s$^{-1}$.
Such an X-ray luminosity seems to be sensibly higher than the X-ray luminosity
of a ``normal'' galaxy. The possibility that the X-ray emission is due to a 
more distant, unidentified object cannot be excluded at present.

\section{Discussion}

In this paper we presented the first detailed X-ray observation 
of the distant Abell cluster A33, obtained with the Beppo-SAX satellite.
We have closely examined and clarified the complex X-ray emission in the
direction of A33.
The analysis of the X-ray data revealed the presence of three different 
X-ray sources in the field of A33. 
The X-ray counterpart of the cluster 
is 1SAXJ0027.2-1930.
We present a spectroscopic redshift for A33,
applying a $\sim 20\%$ correction to the previous photometric estimate.
From optical spectra of six cluster galaxies we measure
a redshift $z=0.2409\pm0.0009$ and a velocity dispersion along the line of
sight $\sigma_{los}$=472$^{+295}_{-148}$ km s$^{-1}$.
The dominant X-ray component (incorrectly
linked with A33 in the past) is associated with a blend
of an AGN and M star, while the X-ray emission
from A33 is $\sim 4$ times fainter.
Using the proper X-ray flux and measured redshift,
we determine a more realistic cluster luminosity of 
$L_{2-10 ~keV} = (7.7\pm0.93) \times 10^{43} ~h^{-2}_{50}$ erg s$^{-1}$, 
one to two orders of magnitude  lower than previous attempts.
The MECS spectral resolution also allows us to determine that 
the  intracluster gas temperature is $T = 2.91^{+1.25}_{-0.54}$
keV. 
No useful information on the cluster abundance is given due to 
the low count rate of the source in the MECS detector.

In the following we will focus on measured quantities 
such as the low temperature and low velocity dispersion. 
We are dealing here with a moderately rich (R=1) and distant 
(D=3) Abell cluster but with X-ray luminosity and temperature more 
typical of nearby (z$ < 0.1$) poor clusters. The temperature of A33 
is commensurate with the predictions from its X-ray 
luminosity from the $L_{X}-T$ relation by David et al. (1993) and 
Arnaud and Evrard (1999). There is an  extensive literature on the 
correlation between these two basic and measurable quantities (Edge \& 
Stewart 1991, Ebeling 1993, David et al. 1993, Fabian et al. 1994, 
Mushotzky \& Scharf 1997, Markevitch 1998, Arnaud and Evrard, 1999).
Comparing the bolometric luminosity of A33 with the best
fit relation, log(L$_{X}$)$=$(2.88$\pm$0.15)
log(T/6keV) $+$(45.06$\pm$0.03) obtained by Arnaud and Evrard (1999),
analyzing a sample of 24 low-z clusters with accurate temperature 
measurements and absence of strong cooling flows, we would expect for 
the A33 a temperature of 3.4 keV, as compared with our deduced value
2.9$^{+1.25}_{-0.54}$. 
The $L_{X}-T$ relation does not  seem to evolve much with 
redshift since z$=$0.4 (Mushotzky \& Scharf 1997). Note however 
that the ASCA data that they use show a strong bias at the low-luminosity
end of the distribution due to the absence of objects in the lower 
luminosity range in the ASCA database. The present data on a cluster at
about 0.2 are thus important to fill in the gap in the $L_X-T$ relationship
found among rich clusters and groups (see Mushotzky \& Scharf 1997). 

The measured velocity dispersion of A33 is also commensurate with 
the predictions from the $\sigma-$T$_{X}$ relationship. 
A large number of authors (see Table 5 in Girardi et al., 1996, or
Table 2 in Wu, Fang and Xu, 1998, for an exhaustive list of
papers on the subject) have attempted to determine the
$\sigma-T$  using different cluster samples in order to test 
the dynamical properties of clusters. 
Girardi et al. (1996) have derived a best fit
relation between the velocity dispersion and the X-ray temperature,
with more than 30$\%$ reduced scatter with respect to previous work
(Edge and Stewart 1991; Lubin and Bahcall 1993; Bird, Mushotzky and
Metzler, 1995; Wu, Fang and Xu 1998, among others).
If we substitute the temperature of 1SAXJ0027.2-1930 in the 
best fit relation log($\sigma$)$=$(2.53$\pm$0.04)+(0.61$\pm$0.05)log(T),
derived by Girardi et al. (1996) a value of 650 km s$^{-1}$
would be expected for the 1-D velocity dispersion, somewhat higher 
but within the uncertainties of the measured value from six 
cluster members of A33.
If we assume energy equipartition between the galaxies and 
the gas in the cluster ($\beta$$=$1) and we use the 
measured temperature of 2.9 keV from the SAX data in the equation 
$\beta = \mu m_p \sigma_v^{2} / k T_{gas} $ (where
$\mu m_p =0.62$, for solar abundance), we  obtain a velocity dispersion 
of $665$ km/s. 

The data for A33 are also consistent with the relation 
$\sigma_{los} \propto (T/keV)^{0.6\pm0.1}$ found by Lubin \& Bahcall (1993) 
and increase its statistical significance in the low temperature 
($T \simlt 3$ keV) range and at intermediate redshifts ($z \sim 0.2$) where 
only a few clusters have measured values of $\beta$. 
This issue will be discussed in a  forthcoming paper. 

We have also found that the bright source 1SAXJ0027.1-1926 has an 
extended appearance which is due to the blending of two different sources: 
an AGN at $z = 0.227$ and approximate  B magnitude $M_B \approx -23.9$ 
(derived from the apparent B magnitude as given in the APM scans) and 
an M-type star. 
The X-ray spectrum does not show any line features, and it is
contaminated by the emission of the M star. Given the
low statistics we did not try to disentangle the two contributions
but we consider an upper limit to the AGN emission using the $F_X/F_V$
for the M star.
The ROSAT BSC source found at a 
position consistent with the coordinates of 1SAXJ0027.1-1926 is 
most probably associated with the AGN. The distance between the foreground 
AGN and the cluster is  $\Delta ~d_L \approx 89.2 h^{-1}_{50} Mpc$. 
At the redshift of the AGN, the observed total flux corresponds to a 
luminosity 
$L_X \simlt 4.5 \times 10^{43}$ erg/s, which can be considered as an upper
limit to the AGN luminosity. 

We also detected a point-like faint source, 1SAXJ0027.0-1928, for which 
no X-ray spectroscopic identification was possible. The $2-10$ keV 
spectrum of this source can be fitted by both thermal
and non-thermal models (see Table 5) but we do not elaborate further given
the poor statistics.

\acknowledgements

S.C. acknowledges useful discussions with  G. Hasinger and C. Sarazin.
Partial financial support from ASI, NASA (NAG5-1880 and NAG5-2523) and
NSF (AST95-00515) grants is gratefully acknowledged.
We appreciate the generosity of B.Tully who allowed us to
take some images and spectra during his observing runs.

\end{document}